\documentclass[12pt]{article}
\begin{document}

\begin{center}
{\Large {\bf Superintegrable Calogero-type systems admit maximal
number of Poisson structures}} \\[10mm] {\Large {\bf C.
Gonera\footnote{Supported by KBN
2 P03B 134 16}}}\\
Department of Field Theory, University of Lodz, Pomorska
149/153,\\ 90-236 Lodz Poland\\ and\\[2mm] {\Large {\bf Y. Nutku}}\\
Feza G{\"u}rsey Institute P. O. Box 6 \c{C}engelk{\"o}y, Istanbul
81220 Turkey\\[3mm] April 25, 2001 \\[10mm]
\end{center}

\noindent {\it We present a general scheme for constructing the
Poisson structure of super-integrable dynamical systems of which
the rational Calogero-Moser system is the most interesting one.
This dynamical system is $2N$ dimensional with $2N- 1$ first
integrals and our construction yields $2N-1$ degenerate Poisson
tensors that each admit $2(N-1)$ Casimirs. Our results are quite
generally applicable to all super-integrable systems and form an
alternative to the traditional bi-Hamiltonian approach.}

\section{Introduction}

Completely integrable systems can be cast into Hamiltonian form in
more than one way \cite{b1}. In fact multi-Hamiltonian structure
can be regarded as an alternative statement of complete
integrability. A Liouville-integrable dynamical system in $2N$
dimensional phase space admits $N$ first integrals which enable us
to construct action-angle variables and reduce the solution to $N$
quadratures. On the other hand there are super-integrable
dynamical systems that admit $2 N - 1$ first integrals so that the
solution requires only one quadrature. It is perhaps worth
underlining that the existence of such additional conserved
Hamiltonians is a rather rare property of dynamical systems and in
no way necessary for Liouville-integrability. An important class
of such super-integrable systems are of Calogero-type \cite{b2}
and its relativistic Ruijsenaars-Schneider generalizations
\cite{b3}, as well as the Winternitz system \cite{b4}. They admit
a complete set, {\it i.e.} maximal number of Poisson structures
but this aspect of super-integrable systems has not received any
attention. In this paper we shall present a general method for the
construction of $2 N - 1$ compatible Poisson structures for
$2N$-dimensional super-integrable dynamical systems. It can be
regarded as a kind of generalization of Nambu mechanics \cite{b5}
and its extensions to integrable dynamical systems with three
degrees of freedom \cite{b6}.

We shall present constructive proof of the complete Poisson
structure of super-integrable systems. To make our
construction explicit, in what follows we shall concentrate only
on the rational Calogero-Moser system but its extension to all
such systems will be manifest. We start with a
definition and brief review of the main properties of this model
in the standard Poisson bracket formulation. Then we point out
that corresponding to the choice of any one of the $2N-1$ first
integrals as the Hamiltonian function there exists a different
Poisson bracket description of motion for the rational
Calogero-Moser dynamics. Each Poisson tensor obtained in this way
is inequivalent to but compatible with all the others. These
Poisson tensors cannot be inverted to yield symplectic structure,
so here we have the example of an even dimensional dynamical
system that admits Poisson but not symplectic structure. We shall
present the explicit form of the complete Poisson structures for
the rational Calogero-Moser system in terms of the original
dynamical variables only for the case of $N=2$ which admits
tri-Poisson structure.

The Hamiltonian structure of the rational Calogero-Moser system
had earlier been discussed by Magri, Morosi and Ragnisco \cite{b7}
and Magri and Mursico \cite{b8}. Their results and those we shall
present in this paper stand in sharp contrast. We present a new
alternative to the bi-Hamiltonian approach of these authors.

\section{Rational Calogero-Moser model}

The $N$-particle rational Calogero-Moser model is defined by the
Hamiltonian function
\begin{equation}
H=\frac{1}{2}\sum_{i=1}^{N}p_{i}^{\;2}+\frac{g^{2}}{2}\sum_{i\neq
j}^N \frac{1}{(q^{i}-q^{j})^{2}}  \label{w1}
\end{equation}
where $g$ is a coupling constant. The canonical coordinates
$q^{i}$ and momenta $p_{j}$ satisfy the standard Poisson bracket
relations
\begin{equation}
\{q^{i},p_{j}\}=\delta_{j}^i \qquad i,j=1,...,N
\label{canonicalpb}
\end{equation}
and we shall find it convenient to refer to the full set of
dynamical variables by a collective symbol
\begin{equation}
x^A \in \left\{ q^1, ..., q^n , p_1, ..., p_n  \right\} \qquad
A=1,...,2N \label{defx}
\end{equation}
so that here and in the following capital Latin indices will range
over $2N$ values. Then
\begin{equation}
\{ x^A, x^B \} = J^{AB}=
\left( \begin{array}{cc}
0&I\\-I&0   \end{array} \right) \label{w8}
\end{equation}
is the canonical Poisson tensor.

It is the existence of a complete set of $N$ functionally
independent and globally defined integrals of motion $I_{k}
\;\;k=1,...,N$ that makes the rational Calogero-Moser model
Liouville integrable. These integrals can be generated from the
conserved quantity $I_{N}$ \cite{b9}
\begin{eqnarray}
I_{N}\equiv exp\left(-\frac{g^{2}}{2}\sum_{i\neq j}
\frac{1}{(q^{i}-q^{j})^{2}} \frac{\partial^{2}}{\partial
p_{i}\partial p_{j}}\right)\prod_{k=1}^{N}p_{k} \label{w2}
\end{eqnarray}
by taking its successive Poisson brackets with $\sum_{i=1}^{N}q^{i}$.
In this way we obtain the conserved Hamiltonians
\begin{eqnarray}
I_{N-n}\equiv \frac{1}{n!}\underbrace{\{\sum_{i=1}^n q^{i}...\{
}_{n-times} \sum_{i=1}^n  q^{i},I_{n}\}...\} \qquad  n=0,1,...,N-1
\label{w3}
\end{eqnarray}
which are in involution with respect to the standard Poisson
bracket defined by (\ref{w8}).

Besides these $N$ first integrals of motion the rational
Calogero-Moser system admits $N-1$ additional functionally
independent integrals of motion $K_{m}$, where $m=2,...,N$ which
are defined by the formula \cite{b10}
\begin{equation}
K_{m}= m g_{1}I_{m}- g_{m} I_{1},  \qquad m=2,...N \label{w4}
\end{equation}
where
\begin{equation}
g_{m} = \frac{1}{2} \, \{\sum_{i=1}^{N} (q^{j})^{\;2}, I_{m} \}
\label{w5}
\end{equation}
with $I_{m}$ given by (\ref{w3}). In general, these super-integrals
are not in involution with respect to the canonical Poisson bracket
defined by (\ref{w8}).

Just as we found it convenient to use a collective label for
coordinates and momenta (\ref{defx}), in what follows it will be
useful to introduce a collective label for the set of first
integrals $I_{n}$ and super-integrals $K_{m}$ according to
\begin{equation}
H_{(\alpha)} \in \{ I_{1},...,I_{n}, K_{2},...,K_{N} \} \qquad
\alpha=1,..,2N-1   \label{w7}
\end{equation}
and use Greek letters exclusively to label the $2N-1$ conserved
Hamiltonians. The parentheses enclosing the Greek indices are
there for a cautionary purpose, namely in the formulae that
follow, they don't take part in the summation convention.

\section{Poisson structures}
\label{sec-ham}

The equations of motion for the rational Calogero-Moser system are
given by
\begin{eqnarray}
\frac{dx^{A}}{d t} \equiv \dot x^{A}=  J^{AB} \partial_{B} H
\equiv J^{AB}\frac{\partial H}{\partial x^{B}} \label{w9}
\end{eqnarray}
with the Poisson tensor (\ref{w8}) and Hamiltonian function (\ref{w1}).
If we consider a trajectory $x^{A}=x^{A}(t)$
which is a solution of equations of motion, then we have
\begin{equation}
\dot{H}_{(\alpha)} = \dot{x}^{K} \partial_{K} H_{(\alpha)}=0 \label{w10}
\end{equation}
and due to the functional independence of the first
integrals $H_{(\alpha)}$, their gradients $\partial_{K}
H_{(\alpha)}$ define $2N-1$ linearly independent vectors
orthogonal to the velocity vector. By taking their full
cross-product we determine a unique direction at each point in
phase space which is precisely that of the velocity vector. Then
the trajectory is given by
\begin{eqnarray}
\dot{x}^{I}=V(x) \; \varepsilon^{IM_{1}...M_{2N-1}} \; (
\partial_{M_{1}} H_{(1)})...(\partial_{M_{2N-1}}H_{(2N-1)})
\label{w11}
\end{eqnarray}
where $\varepsilon^{A_{1}...A_{2N}}$ with $\varepsilon^{1,...,2N} = 1$
is the $2N$-dimensional completely anti-symmetric Levi-Civita symbol
and $V(x)$ is a factor of proportionality which will be determined by the
requirement that both the magnitude as well as the direction of
the velocity vector in phase space is given by eq.(\ref{w11}).
Then the equations of motion will be identical to those of the
rational Calogero-Moser system. This factor of proportionality is
time-independent {\it i.e.} $V$ is a function of integrals of
motion $H_{(\alpha)}$ only and plays the role of volume density in
phase space. So, more properly, the Levi-Civita tensor density is
given by
$$ \epsilon^{A_{1}...A_{2N}} = \frac{1}{V^{-1}} \;
\varepsilon^{A_{1}...A_{2N}} $$
but we shall use it only in the numeric form (\ref{w11}) for ease
of manipulation in the calculations that will follow.

The key to the construction of Poisson tensors associated with
super-integrable dynamical systems such as the rational
Calogero-Moser system is based on the simple observation
that eqs.(\ref{w11}) can be written in the form
\begin{eqnarray}
\dot{x}^{I}=J_{(\alpha)}^{IM_{\alpha}}
\partial_{M_\alpha}H_{(\alpha)} \qquad \alpha=1,...,2N-1
\label{w12}
\end{eqnarray}
where
\begin{equation}
J_{(\alpha)}^{IM_{\alpha}}=V \;
\varepsilon^{IM_{1}...{M_{\alpha}}...M_{2N-1}} (
\partial_{M_{1}} H_{(1)})
...\widehat{(\partial_{M_{\alpha}}H_{(\alpha)}})...
(\partial_{M_{2N-1}}H_{(2N-1)})  \label{w13}
\end{equation}
and a hat over a quantity indicates that it should be omitted.
Therefore, in eqs.(\ref{w13}) we have the expression for $2N-1$
skew-symmetric tensors which yield the equations of motion of the
rational Calogero-Moser system in the Poisson bracket form (\ref{w12}). Of
course to claim that in eqs.(\ref{w13}) we have the definition of
Poisson tensors that define Poisson brackets
$$\{f,g\}_{(\alpha)} \equiv J_{(\alpha)}^{KL} \, (\partial_{K} f) \;
(\partial_{L} g) $$
requires proof that the Jacobi identities are satisfied.
Furthermore if the Jacobi identities are satisfied for a linear
combination of all $2N-1$ skew-symmetric tensors given by
(\ref{w13}), then we have a Poisson pencil.
In order to show that $J_{(\alpha)}^{AB}$ given by eqs.(14)
define Poisson tensors we must prove skew-symmetry and Jacobi
identities
\begin{equation}
J_{(\alpha)}^{K[C}\partial_{K} \; J_{(\alpha)}^{AB]}=0 \label{w16}
\end{equation}
where square brackets denote complete skew-symmetrization. The
former is obvious and in order to prove the latter we
use the identity
\begin{equation}
\varepsilon^{KM_{2}...M_{2N-1}[C}
\varepsilon^{AB]P_{2}...P_{2N-1}} =\frac{2}{(2N)!}
\sum_{i=2}^{2N-1}\varepsilon^{KP_{i}M_{2}...M_{2N-1}}
\varepsilon^{P_{2}...P_{i-1}Cp_{i+1}...P_{2N-1}B A} \label{w17}
\end{equation}
satisfied by the Levi-Civita tensor. This identity is a general
result for the Levi-Civita tensor which is particularly useful for
complete skew-symmetriza\-tion in the indices $A, B, C$ in eq.(\ref{w16}).
There is no particular distinction for different values of $\alpha$ in
eqs.(\ref{w13}) for $J_{\alpha}^{KM}$ and therefore without loss
of generality we can put $\alpha=1$ in proving the Jacobi
identities (\ref{w16}). Then using the definition (\ref{w13}) and
the identity (\ref{w17}) we find that
\begin{eqnarray}
&& \varepsilon^{KM_{2}...M_{2N-1}[C}
\varepsilon^{AB]P_{2}...P_{2N-1}} ( V^{2} \partial_{K} + V
\partial_{K} V) \nonumber \\
& &\Big[ \; (\partial_{M_{2}} H_{(2)})...(\partial_{M_{2N-1}}
H_{(2N-1)}) (\partial_{P_{2}} H_{(2)})... (\partial_{P_{2N-1}}
H_{(2N-1)}) \Big]  \nonumber \\[2mm]
 & = &
\frac{\textstyle{2}}{\textstyle{(2N)!}}\sum_{I=2}^{2N-1}
\varepsilon^{KP_{i}M_{2}...M_{2N-1}}
\varepsilon^{P_{2}...C...P_{2N-1}B A} ( V^{2} \partial_{K} + V
\partial_{K} V) \label{w18}  \\
 & & \Big[ (\partial_{M_{2}}
H_{(2)})...(\partial_{M_{2N-1}}H_{(2N-1)}) (\partial_{P_{2}}
H_{(2)})... (\partial_{P_{2N-1}} H_{(2N-1)}) \Big]  \nonumber
\end{eqnarray}
where the third line above is anti-symmetric in the indices
$P_{i},M_{i}$ but the expression it is contracted with in the last
line above is symmetric in these indices. Thus we have established
that the right hand side of eq.(\ref{w18}) vanishes and
consequently its left hand side, which is just the Jacobi identity
(\ref{w16}), must also vanish.

\subsection{Properties of these Poisson structures}

The compatibility condition for Poisson tensors requires that for
any $\alpha, \;\beta=1,...2N-1$ and $\Lambda \in R$ the linear
combination $\{\cdot\; ,\; \cdot \}_{(\alpha)}+\Lambda\{ \cdot\;,
\cdot\; \}_{(\beta)}$ satisfies the Jacobi identify, thus forming
a Poisson pencil. This compatibility condition is just
\begin{eqnarray}
J_{\alpha}^{K[C}\partial_{K} J_{\beta}^{AB]}+ J_{\beta}^{K[C}
\partial_{K} J_{\alpha}^{AB]} =0 \label{w19}
\end{eqnarray}
and its proof proceeds in the same way as the Jacobi identity.
Using the identity (\ref{w17}) we split (\ref{w19}) into left and
right hand sides as in eq.(\ref{w18}) and show that the right hand
side vanishes.

The general construction of the Poisson tensors for
super-integrable systems that we gave in section \ref{sec-ham}
immediately leads to the fact that they are all degenerate
\begin{equation}
\det | J_{(\alpha)}^{AB} |  =  \varepsilon_{A_1 A_2...A_{2N}}
\varepsilon_{B_1 B_2...B_{2N}} J_{(\alpha)}^{A_1 B_1}
J_{(\alpha)}^{A_2 B_2} ... J_{(\alpha)}^{ A_{2N} B_{2N} } =
 0 \label{degenerate}
\end{equation}
and consequently there exists no symplectic structure corresponding
to any one of the Poisson structures (\ref{w13}).

\subsection{Note added}
\label{sec-referee}

One of the referees of the first version of this paper has pointed
out that we can take the first integrals $H_1, H_2, ..., H_{2N-1}$
together with $H_{2N}$ satisfying
\begin{equation}
\dot{H}_{2N} = 1 \label{one}
\end{equation}
as new coordinates in phase space so that these Poisson structures
assume simply the Darboux form. Then we have
\begin{eqnarray}
\{ H_\alpha, H_{2 N} \}_\beta & = & \delta_{\alpha \beta}
\label{darboux} \\ \{ H_\alpha, H_{\beta} \}_\gamma & = & 0.
\nonumber
\end{eqnarray}
This is always possible locally and in particular we can take
\begin{equation}
H_{2N} = \frac{\Sigma_{i=1}^N \, q^i}{\Sigma_{i=1}^N \, p_i}
\label{2n}
\end{equation}
but there are many other candidates for $H_{2N}$. Here we also see
that $V$ plays the role of the inverse of Jacobian and the fact
that it depends only on $H_{1},...,H_{2N-1}$ can be traced back to
Liouville theorem for original dynamics. The statement
(\ref{darboux}) is a very nice compact way to remember the $2N-1$
Poisson structures.

One can check that this structure coincides locally with the one
given in terms of original phase space coordinates by
eq.(\ref{w12}). However, the equivalence holds only locally as
long as one is able to express original coordinates $x^A$ as
well-defined unique functions of first integrals. But according to
implicit function theorem this is possible only locally. On the
other hand at the heart of the notion of (super-)integrability
lies the requirement of globality, without it any time-independent
Hamiltonian system is, locally, (super-)integrable.

\section{Remarks}

   We have presented a general framework constructing the Poisson
structure of super-integrable systems. It is in a sense an extension of
Nambu's construction of the Poisson structure for the free Euler
top \cite{b5}. We have presented the details for the rational
Calogero-Moser system but it is obvious that this construction can be
repeated for any integrable system which admits the maximal number of
first integrals of motion. There are several important issues raised by
this construction.

        First, the standard Poisson tensor (\ref{w8}) is not
among the $J_{(\alpha)}^{KM}$ structures given by eq.(\ref{w13}).

Each Poisson tensor $J_{(\alpha)}^{AB}$ that generates the
Calogero-Moser dynamics through the Hamiltonian function
$H_{(\alpha)}$ admits $2(N-1)$ Casimirs
\begin{equation}
J_{(\alpha)}^{AB} \; \partial_B H_{(\beta)} =0 \qquad \beta \neq \alpha
\label{casimirs}
\end{equation}
which consist of all the remaining integrals of motion.
Consequently the Calogero-Moser system admits no higher flows and
in this Poisson structure there exists no recursion operator. As a
matter of fact there is no need for the recursion operator because
we start with the full set of integrals of motion from the very
beginning.

Neither one of our Poisson tensors is compatible with the standard
one given by eq.(4).

The Poisson tensors $J_{(\alpha)}^{AB}$ are degenerate for all
$\alpha$ and there exists no symplectic $2$-form.

Finally, we note that we can summarize the whole Poisson structure
of the Calogero-Moser system by writing a single expression
\begin{equation}
 J^{AB} = \sum_{\beta=1}^{2 N-1}
J_{(\beta)}^{AB} \qquad \alpha=1,2,...,2N-1 \label{w22}
\end{equation}
and the equations of motion (\ref{w9}) can be written in the form
\begin{equation}
    \dot{x}^{A}=J^{AB} \partial_{M} H_{(\alpha)}
\end{equation}
for all $\alpha$ because except for the $\alpha^{th}$ term, $H_\alpha$
is a Casimir for each term in the sum (\ref{w22}).

The results we have presented above for the complete Poisson
structure of the general rational Calogero-Moser system stand in
sharp contrast to the bi-Hamiltonian structure of this system
given in \cite{b7} and \cite{b8}. There is an important lesson to
be drawn from this discussion. Namely, there exists alternative
avenues to constructing the Poisson structure of completely
integrable dynamical systems.

\section{Appendix}

We shall now give explicit expressions for our construction of
Poisson tensors in terms of the original canonical coordinates and
momenta for the simplest non-trivial rational Calogero-Moser
system. This is the two particle, $N=2$ system that admits
tri-Poisson structure. First we list the conserved Hamiltonians
\begin{eqnarray}
H_1 & = & p_1 + p_2 \nonumber \\ H_2 & = & \frac{1}{2} (p_1^2 +
p_2^2) +  \frac{g^2}{ 2 (q^1-q^2)^2}  \label{n2hams}\\ H_3 & = & 2
(q^1+q^2) H_2- (p_1 \, q^1 + p_2 \, q^2 ) H_1  \nonumber \\
 &= & p_1^2 \, q_2 + p_2^2 \, q^1 - p_1 p_2 \, (q^1 + q^2)
  + \frac{g^2 ( q^1 + q^2 ) }{(q^1-q^2)^2}
\end{eqnarray}
that will go into the construction (\ref{w13}). Then for the first
Poisson tensor we find
\begin{eqnarray}
J_1^{12} & = & \frac{\textstyle{g^2}}{\textstyle{(q^2-q^1)^3}}\nonumber \\
J_1^{13} & = & \frac{\textstyle{1}}{\textstyle{2 H_2'}}
\left[ ( p_1 - p_2 ) p_1 p_2 + g^2 (p_1+p_2)
\frac{\textstyle{q^1 + q^2}}{\textstyle{(q^1-q^2)^3}} \right] \nonumber\\
J_1^{14} & = & \frac{\textstyle{1}}{\textstyle{2 H_2'}}
\left\{  ( p_1 - p_2 ) p_2^{\;2}
+\frac{\textstyle{g^2}}{\textstyle{(q^1-q^2)^3}} \left[ p_1 ( q^1
+ q^2 ) - p_2 ( q^1 - 3 q^2 ) \right] \right\} \label{j1} \\
J_1^{23} & = & \frac{\textstyle{1}}{\textstyle{2 H_2'}}
\left\{    ( p_2 - p_1 ) p_1^{\;2}
- \frac{\textstyle{g^2}}{\textstyle{(q^1-q^2)^3}}
\left[ p_1 ( 3 q^1 - q^2) + p_2 ( q^1 + q^2 ) \right] \right\} \nonumber
\\
 J_1^{24} & = & \frac{\textstyle{1}}{\textstyle{2 H_2'}}
\left[ ( p_2 - p_1 ) p_1 p_2 - (p_1 + p_2 ) g^2
\frac{\textstyle{q^1 + q^2}}{\textstyle{(q^1-q^2)^3}} \right] \nonumber \\
J_1^{34} & = & \frac{\textstyle{1}}{\textstyle{2 H_2'}}  ( p_1 -p_2)^2
 ( q^1 - q^2 ) \nonumber
\end{eqnarray}
where
\begin{equation}
  H_2' = \frac{1}{2}(p_1 - p_2)^2 +  \frac{g^2}{(q^1-q^2)^2}
 \label{h2p}
\end{equation}
is a conserved quantity and $V^{-1} = 2 H_2'$ serves as the volume
density in phase space. The components of the second Poisson tensor
are given by
\begin{eqnarray}
 J_2^{12} & = & 0 \nonumber \\
 J_2^{13} & = & \frac{\textstyle{1}}{\textstyle{2 H_2'}}  \left[
 ( p_2 - p_1 ) p_1 - g^2 \frac{\textstyle{ 3 q^1 + q^2
}}{\textstyle{(q^1-q^2)^3}} \right] \nonumber  \\
J_2^{14} & = & \frac{\textstyle{1}}{\textstyle{2 H_2'}}  \left[  ( p_2 -
p_1 ) p_2 - g^2 \frac{\textstyle{ q^1 + 3 q^2
}}{\textstyle{(q^1-q^2)^3}}  \right]  \label{j2}
\\J_2^{23} & = & \frac{\textstyle{1}}{\textstyle{2 H_2'}}  \left[
 ( p_1 - p_2 ) p_1 + g^2 \frac{\textstyle{3 q^1 + q^2}}
{\textstyle{(q^1-q^2)^3}}  \right] \nonumber \\
J_2^{24} & = & \frac{\textstyle{1}}{\textstyle{2 H_2'}}  \left[ ( p_1
p_2 ) p_2 + g^2  \frac{\textstyle{q^1 + 3 q^2}}
{\textstyle{(q^1-q^2)^3}} \right] \nonumber \\
J_2^{34} & = & \frac{\textstyle{1}}{\textstyle{2 H_2'}}  \left[
- p_1 ( q^1 + 3 q^2 ) + p_2 ( 3 q^1 + q^2 ) \right] \nonumber
\end{eqnarray}
and the third one is the simplest
\begin{equation}
  J_3 = \frac{1}{2 H_2'} \left( \begin{array}{cccc}
0 & 0 & \frac{\textstyle{g^2}}{\textstyle{(q^1-q^2)^3}} &
\frac{\textstyle{g^2}}{\textstyle{(q^1-q^2)^3}}
\\ 0 & 0 & \frac{\textstyle{g^2}}{\textstyle{(q^2-q^1)^3}} &
\frac{\textstyle{g^2}}{\textstyle{(q^2-q^1)^3}}\\
\frac{\textstyle{g^2}}{\textstyle{(q^2-q^1)^3}} &
\frac{\textstyle{g^2}}{\textstyle{(q^1-q^2)^3}} & 0 & p_1 - p_2
\\ \frac{\textstyle{g^2}}{\textstyle{(q^2-q^1)^3}} &
\frac{\textstyle{g^2}}{\textstyle{(q^1-q^2)^3}} & p_2 - p_1 & 0
\end{array} \right)   .                  \label{j3}
\end{equation}
It can be directly verified that they form a Poisson pencil,
namely
\begin{equation}
 {\cal J} = \sum_{i=1}^3  c_i \, J_i \label{pencil}
\end{equation}
where $c_i$ are arbitrary constants satisfies the Jacobi
identities (\ref{w16}).

\section*{Acknowledgement}

We thank the referee for his remarks reported in section
\ref{sec-referee}. C.G. would like to thank Piotr Kosinski for
very useful comments and discussions.

\end{document}